\documentclass[12pt]{article}
\pdfoutput=1
\usepackage[latin1]{inputenc}

\usepackage{amsmath}
\usepackage{color}
\usepackage{amsfonts}
\usepackage{amssymb}
\usepackage{graphicx}
\usepackage{geometry}
\usepackage{amssymb,epsfig,subfigure}
\usepackage{hyperref}
\usepackage{comment}

\usepackage[numbers,sort&compress]{natbib}

\def\simge{
    \mathrel{\rlap{\raise 0.511ex 
        \hbox{$>$}}{\lower 0.511ex \hbox{$\sim$}}}}

\def\simle{
    \mathrel{\rlap{\raise 0.511ex 
        \hbox{$<$}}{\lower 0.511ex \hbox{$\sim$}}}}

\makeatletter
\renewcommand\section{\@startsection {section}{1}{\z@}%
                                 {-3.5ex \@plus -1ex \@minus -.2ex}
                                   {2.3ex \@plus.2ex}%
                                   {\normalfont\large\bfseries}}
\renewcommand\subsection{\@startsection{subsection}{2}{\z@}%
                                   {-3.25ex\@plus -1ex \@minus -.2ex}%
                                     {1.5ex \@plus .2ex}%
                                     {\normalfont\bfseries}}
\renewcommand\subsubsection{\@startsection{subsubsection}{3}{\z@}%
                                   {-3.25ex\@plus -1ex \@minus -.2ex}%
                                     {1.5ex \@plus .2ex}%
                                     {\normalfont\itshape}}
\makeatother

\def\pplogo{\vbox{\kern-\headheight\kern -29pt
\halign{##&##\hfil\cr&{\ppnumber}\cr\rule{0pt}{2.5ex}&\ppdate\cr}}}
\makeatletter
\def\ps@firstpage{\ps@empty \def\@oddhead{\hss\pplogo}%
  \let\@evenhead\@oddhead 
}
\def\maketitle{\par
 \begingroup
 \def\thefootnote{\fnsymbol{footnote}}
 \def\@makefnmark{\hbox{$^{\@thefnmark}$\hss}}
 \if@twocolumn
 \twocolumn[\@maketitle]
 \else \newpage
 \global\@topnum\z@ \@maketitle \fi\thispagestyle{firstpage}\@thanks
 \endgroup
 \setcounter{footnote}{0}
 \let\maketitle\relax
 \let\@maketitle\relax
 \gdef\@thanks{}\gdef\@author{}\gdef\@title{}\let\thanks\relax}
\makeatother

\numberwithin{equation}{section}

\renewcommand{\dag}{\dagger}

\newcommand{\be}{\begin{equation}}
\newcommand{\bea}{\begin{eqnarray}}
\newcommand{\ee}{\end{equation}}
\newcommand{\eea}{\end{eqnarray}}
\newcommand\beq{\begin{equation}}
\newcommand\eeq{\end{equation}}

\newcommand{\mc}{\mathcal}

\renewcommand{\t}{\tilde}

\def\be{\begin{equation}}
\def\ee{\end{equation}}
\def\ba#1\ea{\begin{align}#1\end{align}}
\def\bg#1\eg{\begin{gather}#1\end{gather}}
\def\bm#1\em{\begin{multline}#1\end{multline}}
\def\bmd#1\emd{\begin{multlined}#1\end{multlined}}

\def\({\left(}
\def\){\right)}
\def\[{\left[}
\def\]{\right]}

\begin{document}

\setcounter{page}0
\def\ppnumber{\vbox{\baselineskip14pt
}}

\def\ppdate{
} \date{}

\author{Shamit Kachru$^{1,2}$, Michael Mulligan$^1$, Gonzalo Torroba$^3$, Huajia Wang$^1$\\
[7mm] \\
{\normalsize \it $^{1}$Stanford Institute for Theoretical Physics, Stanford 
University, Stanford, CA 94305, USA} \\
{\normalsize \it $^2$SLAC National Accelerator Laboratory, 2575 Sand Hill Road, 
Menlo Park, CA 94025, USA }\\
{\normalsize \it $^3$Centro At\'omico Bariloche and CONICET, R8402AGP Bariloche, 
Argentina}
}

\bigskip
\title{\bf  Mirror symmetry and the half-filled \\ Landau level
\vskip 0.5cm}
\maketitle

\begin{abstract}
We study the dynamics of the half-filled zeroth Landau level of Dirac fermions using mirror symmetry, a supersymmetric duality between certain pairs of $2+1$-dimensional theories. 
We show that the half-filled zeroth Landau level of a pair of Dirac fermions is dual to a pair of Fermi surfaces of electrically-neutral composite fermions, coupled to an emergent gauge field. 
Thus, we use supersymmetry to provide a derivation of flux attachment and the emergent Fermi liquid-like state for the lowest Landau level of Dirac fermions.
We find that in the dual theory the Coulomb interaction induces a dynamical exponent $z=2$ for the emergent gauge field, making the interactions classically marginal. 
This enables us to map the problem of $2+1$-dimensional Dirac fermions in a finite transverse magnetic field, interacting via a strong Coulomb interaction, into a perturbatively controlled model. 
We analyze the resulting low-energy theory using the renormalization group and determine the nature of the BCS interaction in the emergent composite Fermi liquid.
\end{abstract}
\bigskip

\newpage

\tableofcontents

\vskip 1cm

\section{Introduction}\label{sec:intro}

Composite fermions provide an intuitive picture for much of the fascinating physics that occurs when strongly interacting electrons are confined to a two-dimensional plane that is pierced by a strong transverse magnetic field \cite{Jain1989, Jainbook, lopezfradkin91, Fradkinbook}.
The dynamics of correlated electrons in a magnetic field is traded for that of composite fermions in reduced effective magnetic flux, interacting through an emergent gauge field.
(See Refs. \cite{zhang1989, Read89, fisher1989} for a closely related theoretical construction in which composite bosons are substituted for the electrons.)
A mean-field treatment of the emergent gauge interaction allows for the interpretation of the gapped abelian fractional quantum Hall states as integer quantum Hall states of composite fermions.

When the inverse electrical filling fraction $\nu_{\rm NR}^{-1}$ (of the assumed spin-polarized non-relativistic electrons), i.e., the ratio of the external magnetic field per flux quantum to the electron density, is an even integer, 
the composite fermions may feel vanishing effective magnetic field within the mean-field approximation and can form a Fermi surface \cite{halperin1993, Kalmeyer1992}.
There is strong experimental evidence for the existence of this metallic phase at half-filling of the lowest and first Landau levels in systems with weak disorder \cite{Jiang1989, willett1990, willett1993, Rokhinson1995, Rokhinson1997, willett1997}; there is evidence for similar gapless states at other even denominator fractions as well \cite{Jiang1992}.
See Ref. \cite{Wong1996} regarding the transition to insulating behavior at stronger values of disorder.
Because a two-dimensional spin-less Fermi liquid is localized in the presence of arbitrarily weak chemical potential disorder \cite{Anderson1979} (and weak external field), this metal cannot be a Fermi liquid \cite{Dobrosavljevic1997, Chakravarty1998}.
Refs. \cite{Manoharan1994, Du1994} provide support for this conclusion by inferring diverging effective masses from magnetoresistance data.

The presence of a Fermi surface of composite fermions in zero effective magnetic field suggests the possibility of a pairing instability, if the pertinent interactions are attractive \cite{greiter1991}.
Indeed, the non-abelian fractional quantum Hall state of Moore and Read \cite{Moore1991} is a candidate state at $\nu_{\rm NR}=5/2$ \cite{willett1987} and can be understood to result from p-wave (angular momentum $l=1$) pairing of the composite fermions \cite{greiter1991, read2000}.
Thus, the composite Fermi liquid represents the gapless parent state from which the most well understood and readily observed (experimentally) examples of systems exhibiting topological order descend \cite{wen1995, nayak2008}.

Because of the attractiveness of this general picture, it is crucial to strengthen our understanding of the duality that relates electrons to composite fermions \cite{shankarmurthyshort, Lee1998, PasquierHaldane, Read1998, simon1998, Stern1999, MurthyShankar2003}.
We shall view the non-relativistic fermions, appropriate for the physics of two-dimensional electronic systems in GaAs heterostructures 
described above, as the low-energy limit of a relativistic system perturbed by symmetry-breaking interactions.
We note that half-filling of the lowest Landau level in a non-relativistic system corresponds to placing the chemical potential at the Dirac point in a relativistic system, i.e., half-filling of the Dirac fermion zeroth Landau level.
Thus, it is worthwhile to first understand the duality for relativistic fermions and this is what we shall do in this paper.
This is interesting given the correlated physics that can occur in graphene \cite{CastroNeto2009} and on the surfaces of time-reversal invariant topological insulators \cite{fu2007a, moore2007,  fu2007b, qi2008b, roy2009a}. 

The duality that we study was proposed in a supersymmetric context by Intriligator and Seiberg~\cite{Intriligator:1996ex} (with
various extensions appearing in ~\cite{deBoer:1996mp, deBoeroz,deBoeroztwo,Aharony}). 
A useful interpretation of the equivalence in terms of dual partition functions was provided by Kapustin and Strassler \cite{Kapustin:1999ha}.
A proof of the duality was essentially given by Borokhov, Kapustin, and Wu \cite{BorokhovKapustinWumirrormonopoles} by matching the Hilbert space of the two theories.
Additional evidence for the duality was provided recently by a matching of 3-sphere partition functions by Kapustin, Willett, and Yaakov \cite{KapustinWillettYaakov}.
Following convention, we refer to the duality as mirror symmetry.
As we review in the next section, mirror symmetry provides dual descriptions of equivalent physics.
In the absence of symmetry-breaking perturbations, it relates two supersymmetric field theories to one another.
We refer to these theories as theory A, which is conventionally called the ``magnetic theory," and theory B, conventionally called the ``electric theory."

In this work, we will deform mirror symmetry by the addition of an external magnetic field to theory A. (The study of mirror symmetry in the presence of external sources was initiated in~\cite{Hook:2013yda, Hook:2014dfa}). 
A background magnetic field in theory A breaks supersymmetry and reduces the effective low-energy physics to that of two flavors of Dirac fermions at zero density, i.e., half-filling of the zeroth Landau level.
The bosonic superpartners are effectively projected out of the low-energy physics in the limit of zero Landau level mixing, i.e., as $B \rightarrow \infty$.\footnote{Corrections due to Landau level mixing may be numerically small even for finite $B$ \cite{BisharaNayak}.}
The removal of the bosonic superpartners has the benefit of reducing the physics to that which is closely related to systems that can be realized experimentally.
We note, however, that these two Dirac fermions carry opposite charge under the $U(1)_J \equiv U(1)_{\rm EM}$ symmetry that we identify with electromagnetism and so the exact low-energy theory that we study cannot be strictly identified with those found in an actual experimental system.

The magnetic field in theory A translates into a non-zero density of composite fermions in theory B.
Again, this breaks the supersymmetry of theory B.
The bosonic superpartners, along with the additional matter fields, of theory B are removed in the strong coupling limit of interest, as we shall explain, thereby leaving us with the physics of composite Dirac fermions at finite density interacting through an emergent gauge field at low energies. 
The Dirac fermions of theory B are electrically neutral. This provides a dynamical derivation of flux attachment and the composite Fermi liquid picture in the present model.

Fluctuations of the electromagnetic gauge field are naturally traced through the duality.
These interactions have two important effects which can be clearly seen in theory B.
First, the fluctuating electromagnetic gauge field strongly modifies the dispersion of the emergent gauge field of theory B and renders the interaction between the emergent gauge field and the composite fermions marginal.
In particular, a strong Coulomb interaction in theory A is mapped into a perturbatively small fermion--boson coupling in theory B.
Second, this fluctuating Coulomb field mediates a BCS pairing interaction that we find to be repulsive. 
Therefore, at least away from asymptotically low energies, the emergent composite Fermi liquid is stable to pairing; additional interactions are necessary to induce attraction.
Thus, we are able to understand the low-energy dynamics of the strongly interacting quantum Hall system using our weakly-coupled dual.

The remainder of this paper is organized as follows.
In \S2, we review the relevant aspects of mirror symmetry.
In \S3, we deform mirror symmetry by the addition of a magnetic field, and describe its application to the half-filled zeroth Landau level problem of Dirac fermions.
In \S4, we study the low-energy dynamics of theory B, set up a renormalization group analysis, and determine the nature of the superconducting interaction.
We conclude in \S5 and provide an Appendix that elaborates upon some aspects of the formalism used in the main text.

\section{Mirror symmetry in $2+1$ dimensions}\label{sec:overview}

We begin by reviewing mirror symmetry for $D=2+1$ dimensional supersymmetric theories~\cite{Intriligator:1996ex, deBoer:1996mp, Kapustin:1999ha}. For our purpose, it will be sufficient to consider the simplest mirror pair, namely $U(1)$ supersymmetric QED (SQED) with one flavor, and the theory of a free hypermultiplet; we follow mostly the analysis of Kapustin and Strassler~\cite{Kapustin:1999ha}.
We then describe how to include electromagnetism.


\subsection{Theory A}

``Theory A'' (sometimes also called the magnetic theory), which will be identified below with the elementary electrons of the quantum Hall system, is the theory of a free hypermultiplet with $\mc N=4$ supersymmetry (i.e., 8 supercharges).
Each conserved supercharge is a fermionic operator that commutes with the Hamiltonian and together generate the supersymmetry algebra.  
In $\mc N=2$ notation, this is given by two chiral multiplets $(V_+, V_-)$, each of which contains a complex scalar $v_\pm$ and a 3D Dirac fermion $\Psi_\pm$. A crucial role will be played by a $U(1)_J$ global symmetry, under which the supermultiplets $V_\pm$ have charges $\pm 1$. This symmetry will be identified with $3+1$ dimensional electromagnetism. The theory has nonabelian $SU(2)_L \times SU(2)_R$ R-symmetries, under which $(v_+, v_-^*)$ and $(\Psi_+, \Psi_-^*)$ transform as $(2,1)$ and $(1,2)$, respectively. This is summarized in (\ref{tab:thA}).
\begin{center}
\be\label{tab:thA}
\begin{tabular}{c|ccc}
&$SU(2)_L$&$SU(2)_R$&$U(1)_J$\\
\hline
&&&\\[-12pt]
$(v_+, v_-^*)$  & $2$ & 1 & 1  \\
&&&\\[-12pt]
$(\Psi_+, \Psi_-^*)$  & 1& $2$ & 1
\end{tabular}
\ee
\end{center}

The Lagrangian is simply that of free fields,
\be\label{eq:mag1}
L^{\rm (A)} = \sum_{\pm} \Big(|\partial_\mu v_\pm|^2+i \bar \Psi_\pm \not \! \partial \Psi_\pm\Big)\,,
\ee
where $\bar{\Psi}_{\pm} = \Psi^\dagger \gamma^0$.
We work in metric signature $(+--)$, and a convenient choice for gamma matrices is
\be\label{eq:gammas}
\gamma^0=\sigma_3\,,\,\gamma^1=i\sigma_1\,,\,\gamma^2=i\sigma_2\,.
\ee
The current of the $U(1)_J \equiv U(1)_{\rm EM}$ symmetry is given by
\be\label{eq:Jmu1}
J_\mu =\sum_\pm q_\pm \,\left(\bar \Psi_\pm \gamma_\mu \Psi_\pm+iv_\pm \partial_\mu v_\pm^*-i v_\pm^*\partial_\mu v_\pm \right)\,,
\ee
where $q_{\pm} = \pm 1$. 
We will shortly add an external magnetic field and charge density.

\subsection{Theory B}

``Theory B'' (a.k.a., the electric theory) is $D=2+1$ dimensional SQED with $\mc N=4$ supersymmetry and a charged hypermultiplet. It will provide a concrete realization of duality and flux attachment for the quantum Hall fluid, in terms of composite fermions and an emergent gauge field. The $\mc N=4$ vector multiplet contains an $\mc N=2$ vector multiplet $V=(a_\mu, \sigma, \lambda)$ and an $\mc N=2$ neutral chiral multiplet $\Phi= (\phi, \psi_\phi)$. Here $\sigma$ is a real scalar, $\phi$ is a complex scalar, and $\lambda$ and $\psi_\phi$ are Dirac fermions. The $\mc N=4$ charged hypermultiplet contains $\mc N =2$ chiral multiplets of opposite charge, $Q_\pm = (u_\pm, \psi_\pm)$. The fermions $\psi_\pm$ will play the role of composite fermions in the QHE.

The $U(1)_J \equiv U(1)_{\rm EM}$ global symmetry of theory B arises from dualizing the field strength,
\be\label{eq:Jmu2}
J_\mu=\frac{1}{2\pi} \epsilon_{\mu\nu\rho} \partial^\nu a^\rho\,,
\ee
whose conservation law is equivalent to the Bianchi identity for the emergent gauge field. It acts as a shift on the dual photon $\gamma$, where $f_{\mu\nu}=\partial_\mu a_\nu-\partial_\nu a_\mu=\epsilon_{\mu\nu\rho} \partial^\rho \gamma$. In the duality, the $U(1)_J$ global symmetries of both theories are identified.
The gauge field then arises from dualizing the matter current (\ref{eq:Jmu1}) of theory A.
The rest of the fields are {\it neutral} under $U(1)_J \equiv U(1)_{\rm EM}$. On the other hand, the symmetries $SU(2)_L \times SU(2)_R$ act as $(3,1)$ on the triplet of scalars $(\sigma, \phi)$ (recall that $\sigma$ is real and $\phi$ is complex), $\lambda, \psi_\phi$ are in the bifundamental, $(u_+, u_-^*)$ transform as $(1,2)$, and $(\psi_+, \psi_-^*)$ are in the $(2,1)$. This is summarized in (\ref{tab:thB}).
\begin{center}
\be\label{tab:thB}
\begin{tabular}{c|ccc}
&$SU(2)_L$&$SU(2)_R$&$U(1)_J$\\
\hline
&&&\\[-12pt]
$e^{2\pi i \gamma/g^2}$  & 1& 1 & 1  \\
&&&\\[-12pt]
$\phi_{ij}=(\sigma, \phi)$  & 3& 1 & 0  \\
&&&\\[-12pt]
$\lambda_{ia}=(\lambda, \psi_\phi)$  & 2& 2 &  0 \\
&&&\\[-12pt]
$u_a=(u_+, u_-^*)$  & 1& 2 & 0  \\
&&&\\[-12pt]
$\psi_i=(\psi_+, \psi_-^*)$  & 2& 1 & 0 
\end{tabular}
\ee
\end{center}

The Lagrangian of theory B is fixed by the symmetries and is nontrivial due to the interactions between the charged hypermultiplet and the emergent vector multiplet:
\be\label{eq:electric1}
L^\text{(B)}= L_V(\mathcal V) + L_H(\mathcal Q, \mathcal V),
\ee
where the kinetic terms for the vector superfield are
\be\label{eq:LV}
L_V(\mathcal V) =\frac{1}{g^2} \left(- \frac{1}{4} f_{\mu\nu}^2 + \frac{1}{2} (\partial_\mu \phi_{ij})^2+ i \bar \lambda_{ia} \not \! \partial \lambda_{ia}+ \frac{1}{2} D_{(ab)}^2 \right)
\ee
and the hypermultiplet part of the Lagrangian reads
\be
L_H(\mathcal Q, \mathcal V)= |D_\mu u_a|^2 + i \bar \psi_i \not \! \!D \psi_i- \phi_{ij}^2 |u_a|^2- \phi_{ij} \bar \psi_i \psi_j + \sqrt{2}(i\lambda_{ia} u^\ast_a \psi_i + {\rm h.c.})+ D_{(ab)} u_a^\ast u_b\,.
\ee
Here $D_\mu= \partial_\mu+i q_{\pm} a_\mu$ and $D_{(ab)}$ are the auxiliary fields from the vector multiplet; integrating them out leads to a quartic potential $V=\frac{g^2}{2} (u^\ast_a u_b)^2$ for the hypermultiplet scalars.

\subsection{Mirror symmetry}

In $2+1$ dimensions the gauge interaction is classically relevant; as a result, theory B flows to strong coupling at low energies. Mirror symmetry states that the low energy limit of theory B admits a dual description as the model of a free hypermultiplet given by theory A. This can be proved by a formal path integral calculation in the limit $g^2 \gg E$~\cite{Kapustin:1999ha}. More physically, theory A arises as the low energy description of theory B along the ``Coulomb branch" of its moduli space where the emergent gauge field is deconfined; the power of supersymmetry here is that such an effective theory is one loop exact -- both perturbatively and nonperturbatively. 

Theory A has a ``Higgs branch'',\footnote{This nomenclature is related to the fact that in generalizations of mirror symmetry to many flavors, this is a branch along which gauge symmetries are spontaneously broken.} a moduli space of vacua parametrized by the complex fields $v_\pm$. Such moduli spaces are protected by supersymmetry, but will be shortly lifted by the addition of a magnetic field to realize the Landau levels. On the other hand, theory B has a Coulomb branch where the triplet of scalars $\phi_{ij}$, together with the dual photon $\gamma$, have nonzero expectation values. Along these directions, the $U(1)$ gauge symmetry is preserved, and the charged hypermultiplet fields become massive.
The duality maps the Coulomb branch of theory B to the Higgs branch of theory A; note that there is no Higgs branch for theory B due to the constraints $D_{ab}=0$ which give the absolute minimum of the potential. 

An explicit derivation of theory A from theory B may be obtained as follows \cite{SachdevYin}.
Away from the origin $\phi_{ij} = 0$ of the Coulomb branch, we may integrate out the heavy hypermultiplets of theory B to obtain a nonlinear sigma model for $(\phi_{ij}, \gamma)$. Due to nonrenormalization theorems of supersymmetry, and the absence of nonperturbative effects for $U(1)$ SQED, this model is exact at one loop. Theory A can then
be obtained explicitly by taking the low energy limit of the nonlinear sigma model along the Coulomb branch of theory B.
This gives, in the low energy limit $|\phi |/g^2 \ll 1$,
\be\label{eq:quantum-mapb}
v_i \equiv \left(
\begin{matrix}
v_+ \\ v_-^*
\end{matrix}\right)= \sqrt{\frac{|\vec \phi|}{2\pi}} e^{2\pi i \gamma/g^2}
\left(
\begin{matrix}
\cos \frac{\theta}{2} \\
e^{i \varphi} \sin \frac{\theta}{2}
\end{matrix}
\right)\;\;,\;\;\vec \phi = |\vec \phi| (\cos \theta, \sin \theta \cos \varphi, \sin \theta \sin \varphi),
\ee
and for the fermions
\be\label{eq:quantum-mapf}
\Psi_a \equiv \left(
\begin{matrix}
\Psi_+ \\ \Psi_-^*
\end{matrix}\right)= \frac{1}{\sqrt{2}}\,\frac{\lambda_{ai} v_i }{2\pi\,\sum_i|v_i|^2}\,.
\ee
More details of the duality, with possible other applications to phenomena in condensed 
matter physics, were recently reviewed in~\cite{Hook:2013yda, Hook:2014dfa} (building on 
earlier work of \cite{SachdevYin}, which also explored connections to modern ideas in condensed matter physics).
Remarkably, the duality continues to hold even at the origin of the Coulomb branch when $\phi_{ij} = 0$ where theory B is a strongly interacting conformal field theory.
For this case, monopole operators of theory B can be obtained and identified with the free fields of theory A \cite{BorokhovKapustinWumirrormonopoles}.

This is the content of mirror symmetry for this pair of theories. Even though we have presented this as an infrared duality (theory B flows to theory A at energies much smaller than $g^2$), the correspondence can in fact be extended to all energy scales. In this case, theory A is deformed by irrelevant operators in powers of $E/g^2$ that encode the nontrivial sigma model along the Coulomb branch of theory B (the so-called Taub-NUT geometry).

In this duality there is a natural direction for the RG, with theory B providing the weakly coupled UV fixed point, and theory A emerging as the low energy description at scales $E \ll g^2$. However, we will find that in the presence of an external magnetic field both descriptions turn out to be useful in the IR. In particular, theory A will provide the elementary electrons of the quantum Hall system and theory B will display a Fermi surface of composite fermions coupled to an emergent gauge field $a_\mu$. In fact, after turning on Coulomb interactions with strength $e$, we will derive a new strong/weak duality with $e \to 1/e$.

\subsection{Adding electromagnetism}

To proceed, we will weakly gauge the global $U(1)_J \equiv U(1)_{\rm EM}$ symmetry and identify it with electromagnetism in the quantum Hall system of interest. In theory A, this adds a gauge field $A_\mu$ with a 4D kinetic term, so that the action becomes
\be\label{eq:mag2}
S^{\rm (A)}=\sum_\pm \int d^3x\Big\{ |(\partial_\mu + i  q_\pm A_\mu) v_\pm|^2+ \bar \Psi_\pm \gamma^\mu(i\partial_\mu -   q_\pm A_\mu) \Psi_\pm\Big\}- \frac{1}{4e^2} \int d^4x F_{\mu\nu}^2\,.
\ee
Again, $q_\pm= \pm 1$, and $e$ is the 4D electromagnetic coupling. From the viewpoint of the 3D theory, $A_\mu$ behaves as a background gauge field.

The works~\cite{Hook:2013yda, Hook:2014dfa} studied in detail the effects of adding external sources to mirror symmetry, and our analysis here will follow similar steps. The main point is that the $U(1)_J$ background gauge field, under which the free hypermultiplet fields of theory A carry elementary charge, appears as a BF interaction involving the emergent gauge field of theory B. As a result, after weakly gauging the $U(1)_J$ and allowing for a background $A_\mu$, the action for theory B becomes
\be\label{eq:electric2}
S^{\rm (B)} =\int d^3x \,\left\{\sum_\pm \bar \psi_\pm \gamma^\mu(i\partial_\mu- q_\pm a_\mu) \psi_\pm - \frac{1}{4\pi} \epsilon^{\mu\nu\rho} a_\mu F_{\nu\rho}+\ldots\right\}- \frac{1}{4e^2} \int d^4x F_{\mu\nu}^2\,.
\ee
where `$\ldots$' are the additional terms shown in (\ref{eq:electric1}).
Indeed, recalling the identification of $U(1)_J$ currents discussed before, the BF term is the same as the coupling between the matter current and the external gauge field in theory A,
\be\label{eq:BFdual}
 - \frac{1}{4\pi} \int d^3x\,\epsilon^{\mu\nu\rho} a_\mu F_{\nu\rho}= \int d^3 x J_\mu A^\mu\,.
\ee
The coefficient in the BF action is fixed by the normalization of $a_\mu$ chosen in (\ref{eq:Jmu2}). 

We will find that this correspondence is essential for the derivation of the duality between Dirac fermions at $\nu=1/2$ and the composite fermion model. Eq.~(\ref{eq:BFdual}) will implement flux attachment in a dynamical and adiabatic way.

\section{Duality for the half-filled zeroth Landau level}\label{sec:duality}

We are now ready to analyze the half-filled zeroth Landau level for Dirac fermions. For this, we will go beyond the relativistic mirror symmetry of \S \ref{sec:overview} and turn on a magnetic field and a finite density of fermions. Our strategy will be to derive first the most important features of the duality and exhibit how flux attachment works with the BF coupling (\ref{eq:BFdual}). We will do this by working in a formal limit $g^2 \to \infty$ in theory B, which sets to zero the kinetic terms for the gauge field and its superpartners in (\ref{eq:LV}).\footnote{This approach was also used for the proof of mirror symmetry in~\cite{Kapustin:1999ha}.} Imposing the equations of motion of the constraint fields will then reveal very simply the appearance of a Fermi surface for composite fermions. Next in \S \ref{sec:dynamics} we will take into account finite $g$ effects and will analyze the low-energy dynamics for the Fermi surface excitations interacting with the emergent gauge field (and its superpartners).

\subsection{The quantum Hall system}

Let us first turn on a magnetic field for electromagnetism (the $U(1)_J$ symmetry) in theory A of \S \ref{sec:overview},
\be
B= \frac{1}{2} \epsilon_{zij} F_{ij}\,.
\ee
From the action (\ref{eq:mag2}), the Landau levels for the fermions and scalars are
\be
E_\text{fermion}^{(n)}= \pm \sqrt{2n B}\;,\;E_\text{boson}^{(n)}= \pm \sqrt{(2n+1) B}\,.
\ee
Importantly for our purpose, the fermions have a zeroth Landau level with $E=0$, but the scalars are gapped. The low energy dynamics in the zeroth Landau level will then be dominated by the two oppositely charged Dirac fermions, with no traces from the scalars required by supersymmetry. In this way, even though we start from a supersymmetric system, after turning on the magnetic field (which breaks SUSY explicitly) the resulting theory has no light scalars. Already at the classical level, the scalars are stabilized at the origin,
\be\label{eq:originA}
\langle v_\pm \rangle =0\,.
\ee

Projecting down to the zeroth Landau level, the matter content of the theory is then simply that of two oppositely charged relativistic Dirac fermions. This field content is related to systems such as graphene (where, in distinction, the charges of the electrons are the same), and also as a ``parent'' theory from which non-relativistic fermions can be obtained by additional deformations -- a point which we plan to analyze in future work. It is necessary to stress also the appearance of an even number of fermions, making the theory regularizable (in a symmetry-preserving manner at weak coupling) and, thus, free of the parity anomaly \cite{NiemiRQ, RedlichKN, RedlichDV}. 
The even number of fermions will also appear below in a slightly different way in theory B. 
This is an important difference with recent works \cite{Metlitski:2015eka, 2015arXiv150505141W} that study duality in a system that features a single Dirac fermion in 2+1 dimensions and are, therefore, realized (at weak coupling) on the boundary of a 3+1 dimensional spacetime in order to escape the parity anomaly constraint \cite{MulliganBurnell2013}. 

Consider moving away from the zero density Dirac point by turning on a $U(1)_J \equiv U(1)_{\rm EM}$ chemical potential $\mu_F = \langle A_0 \rangle$.
The filling fraction of the Dirac fermions,
\be\label{eq:filling1}
\nu = \frac{n}{B/2\pi}\,,
\ee
where $n$ is the charge density, proportional to $\mu_F^2$.  
When the chemical potential is placed at the origin, $\mu_F = 0$, the zeroth Landau level is said to be half-filled.
For simplicity, we will focus on configurations with vanishing $SU(2)_R$ charge, for which the $\Psi_+$ and $\Psi_-^*$ levels are equally occupied.\footnote{An imbalance in this charge may also be interesting, being dual to turning on expectation values for $u_\pm$.} 
We are interested in the regime $\nu \ll 1$, which corresponds to filling fractions $\nu_{NR} \sim 1/2$ for each species of non-relativistic fermion. 

In the simplest version of the duality, $\mu_F \ll B^{1/2}$ is required so that the scalars $v_\pm$ remain gapped. Indeed, the $v_\pm$ may become unstable as $\nu \sim 1$, because the chemical potential tends to induce a Bose-Einstein condensate. Fortunately, this is not the regime of interest in this work. For future reference, we point out that it is possible to turn on additional supersymmetry breaking deformations to stabilize $v_\pm$ against the chemical potential, and map them through the duality using the explicit dictionary (\ref{eq:quantum-mapb}). In this way, it may be possible to study larger filling fractions without the danger of runaways from the elementary scalars.

\subsection{Dual description}\label{subsec:dualdescription}

By mirror symmetry, Dirac fermions, each at filling fraction $\nu \ll 1$, admit a dual description where the background magnetic field and chemical potential couple to the emergent gauge field via the BF interaction in (\ref{eq:electric2}). 
For the duality to be valid, it is necessary to choose $\mu_F$ and $B^{1/2}$ much smaller than $g^2$, the energy scale set by the square of the emergent gauge field coupling constant.

Let us analyze the dynamics at low energies $E \ll g^2$.
We take the formal limit $g \to \infty$, where the kinetic term for the vector multiplet goes to zero, as explained in~\cite{Kapustin:1999ha}.  Focusing first on the hypermultiplet scalars, the F-term and D-term equations of motion set
\be
u_+ u_-=0, \quad |u_+|^2 = |u_-|^2\,,
\ee
and require: $\langle u_+ \rangle = \langle u_- \rangle =0$. 
This is the familiar fact that the SQED theory with $N_f=1$ hypermultiplets has no Higgs branch, as it is lifted by the D and F-term conditions. 
This conclusion also follows from the absence of $SU(2)_L$ symmetry breaking in theory A.
Note that at finite but large $g$ this constraint is implemented in a smooth way by a potential $V_D= \frac{g^2}{2} (u^\ast_a u_a)^2$. 
This constraint allows us to evade the intuition that a finite density of fermions is necessarily accompanied by a finite density of bosons in a perturbed supersymmetric field theory.

Consider next the hypermultiplet fermions. 
Recall our convention of denoting $\psi_i = (\psi_+, \psi_-^*)$ and our gamma matrix choice in (\ref{eq:gammas}).
The relevant equations of motion come from extremizing with respect to $\phi_{ij}$ and $a_\mu$, which give, respectively,
\be\label{eq:cond1}
\langle \bar \psi_i \psi_j\rangle =0
\ee
and
\be\label{eq:cond2}
\langle\bar \psi_i \gamma_\mu \psi_i \rangle= -\frac{1}{4\pi} \epsilon_{\mu\nu\rho} F_{\nu\rho}\,.
\ee
The $a_0$ equation is the Gauss' law constraint, imposing charge neutrality for the $U(1)$ gauge symmetry: a finite density of composite fermions cancels the background contribution from $F_{\mu\nu}$.

Furthermore, in theory A the $SU(2)_R$ charge of the vacuum vanishes, which translates in theory B into
\be\label{eq:cond3}
\langle \psi_i^\dag T_{ij}^A \psi_j \rangle =0\,,
\ee
with $T^A$ the $SU(2)$ generators. The vanishing of this charge in theory A is seen classically from the stabilization of $v_\pm$ at nonzero magnetic field. We expect that the dual statement (\ref{eq:cond3}) should be seen quantum-mechanically; here for simplicity we impose this equation from the beginning.

Eq.~(\ref{eq:cond2}) implies that the magnetic field in the original system becomes a density of the composite fermions $\psi_i$. As a result we obtain an emergent Fermi surface with charge fixed by the magnetic field:
\be
\langle \psi^\dag_+ \psi_+ \rangle-\langle \psi^\dag_- \psi_- \rangle=-\frac{B}{2\pi}\,.
\ee
We have to ensure that the Fermi surface can be filled consistently with the constraints (\ref{eq:cond1}) and (\ref{eq:cond3}). 

Let us denote the two components of a single Dirac fermion by $\psi_{\pm}=(\psi_{\pm, \uparrow}, \psi_{\pm, \downarrow})$.
First, (\ref{eq:cond1}) with $i \neq j$ automatically vanishes on the Fermi surface vacuum, because the two fields create and destroy different types of fermions. The $i=j$ conditions imply, 
\be
\langle \psi_{+, \uparrow}^\ast \psi_{+, \uparrow} \rangle-\langle \psi_{+, \downarrow}^\ast \psi_{+, \downarrow} \rangle=\langle \psi_{-, \uparrow}^\ast \psi_{-, \uparrow} \rangle-\langle \psi_{-, \downarrow}^\ast \psi_{-, \downarrow} \rangle=0\,,
\ee
so that the spin up and spin down fermions for each fermion flavor are filled symmetrically.
On the other hand, the condition (\ref{eq:cond3}) that the $SU(2)_L$ current $J_0^3$ vanishes requires that the Fermi surfaces of the two flavors of fermions are equally filled,
\be
\langle \psi_+^\dag \psi_+ \rangle =-\langle \psi_-^\dag \psi_- \rangle\,.
\ee
The currents $J_0^\pm$ vanish trivially on the Fermi surface vacuum because, again, they are bilinears in fermions with different flavors.

In this way, Dirac fermions in 2+1 dimensions in the presence of a magnetic field at zero density are dual to a Fermi liquid of composite fermions (neutral under electromagnetism) interacting with an emergent gauge field and its superpartners. 
For $B>0$, we obtain a Fermi surface of antiparticles of $\psi_+$ fermions and a Fermi surface of particles of $\psi_-$ fermions, and vice versa for $B<0$.
Thus, time-reversal, which maps the applied magnetic field $B \rightarrow - B$ of theory A, switches the sign of the induced chemical potential of theory B.\footnote{See Ref. \cite{Son:2015xqa} for further details on the mapping of discrete symmetries across the duality.}
This provides a derivation of the duality between the half-filled zeroth Landau level (for each species of fermion) and a composite Fermi liquid in the context of a particular UV completion. Flux attachment happens dynamically and adiabatically as a function of $g^2$. It does not occur due to a Chern-Simons term for $a_\mu$ (here forbidden by symmetries), but rather it happens through the BF coupling (\ref{eq:electric2}), as we just found. Furthermore, a chemical potential $\mu_F$ in the original (electronic) system gives a nonzero filling fraction $\nu$, Eq. (\ref{eq:filling1}), and in the dual description it becomes a density of magnetic monopoles. This was studied in~\cite{Hook:2013yda}.

This constitutes a full dynamical duality between the quantum Hall system and the emergent non-Fermi liquid. 
The proposed dual of theory B has a specific matter content and interactions, and is a consequence of mirror symmetry. We note the presence of the gauginos $\lambda_{ia}$ in the low-energy effective theory.
At vanishing external field $B=0$ and away from the origin $\phi_{ij} = 0$, we reviewed in (\ref{eq:quantum-mapf}) how the gauginos bind with the dual photon and give rise to the elementary excitations of theory A.
We expect the $B \neq 0$ case to be analogous to the case with $\phi_{ij} = 0$ at vanishing external field in the sense that an identification of the theory A excitations proceeds through a careful study of the monopole operators of theory B.
Note also that there is an even number of composite fermions, making the theory regularizable in a manner that preserves the discrete symmetries of the IR theory and, thus, free of the parity anomaly \cite{NiemiRQ, RedlichKN, RedlichDV}. 

In the recent paper~\cite{Son:2015xqa} that partially motivated our work, Son proposed that the half-filled zeroth Landau level for a Dirac fermion has a dual in terms of a single composite fermion interacting with an emergent gauge field, which in turn has a BF coupling of the type we found above. These elements appear in our proposal as well. This was also studied from the point of view of three-dimensional topological insulators in ~\cite{Metlitski:2015eka, 2015arXiv150505141W}. Such constructions could not establish a dynamical duality between the quantum Hall and Fermi surface systems, something that here we find as a consequence of supersymmetry. In fact, as we discussed in \S \ref{sec:overview}, the quantum Hall system emerges as the effective nonlinear sigma model description along the Coulomb branch (expectation values of the dual photon and its partners) of the theory for the Fermi surface excitations interacting with the emergent gauge field. In the remainder of the work we will study the consequences of this duality.

\subsection{Mapping of filling fractions}\label{subsec:filling}

Having established the exact mapping between global quantities across the duality, we now use it to compute the mapping of filling fractions near half-filling of the zeroth Landau level by starting from theory A (in the UV), in a state with a small total charge density $J_0$ in a strong magnetic field $B$. 
The bosons have been gapped out. What remains are two copies of charged fermions $(\Psi_+, \Psi_-^*)$ of equal charge, with individual charge densities $J_{\pm}$, and total electrical charge $J_0=J_+ +J_-$.  Notice that $(\Psi_+, \Psi_-^*)$ also carry opposite charges under the Cartan subgroup\footnote{That is, the maximal commuting subgroup which for $SU(2)$ is $U(1)$.} 
of $SU(2)_R$, whose charge density is given by $J_0^R=J_{+}-J_{-}$. In other words, preservation of the global symmetry $SU(2)_R$ enforces equal charge densities between the two fermions: $J_+=J_-$. 
Each individual Dirac fermion has filling fraction,
\be
2\;\nu_{\pm}=\frac{J_0}{B/2\pi}\,.
\ee  

Now we go to the dual description: theory B in the formal $g\rightarrow \infty$ limit. There are two electrically neutral fermions $(\psi_+,\psi_-^*)$ charged equally under the emergent gauge field $a_\mu$. The global quantities $(J_0, B)$ are mapped over via the BF coupling:
\bea
L_{BF}&=&-\frac{1}{4\pi}\epsilon^{\mu\nu\rho} a_\mu F_{\nu\rho}=-\frac{1}{2\pi} a_0 \left(\frac{1}{2}\epsilon^{ij}F_{ij}\right) + \frac{1}{2\pi}A_0 \left(\frac{1}{2}\epsilon^{ij}f_{ij}\right)\nonumber\\
&=& -\frac{1}{2\pi}a_0\;B +\frac{1}{2\pi}A_0 \tilde{B}\,,
\eea
where $\tilde{B}$ is the magnetic field of the emergent gauge field.

Taking the equation of motion with respect to $a_0$ gives the charge density under the emergent gauge symmetry:
\be
\tilde{J}_0 = \tilde{J}_++\tilde{J}_-=-\frac{B}{2\pi}\,.
\ee
In particular, $(\psi_+,\psi_-^*)$ carry opposite charges under the Cartan of $SU(2)_L$ symmetry that is also present in theory A. Neutrality under this global symmetry enforces $\tilde{J}_+=\tilde{J}_-=\tilde{J}_0/2$.

The charge density of the original $U(1)_{\rm EM}$ appears in the dual theory as:
\be
J_0 = \frac{\delta S}{\delta A_0}=\frac{\tilde{B}}{2\pi} \,.
\ee
Therefore in theory B we have two copies of fermions $(\psi_+,\psi_-^*)$, whose charge densities under the emergent $a_\mu$ equally split $\tilde{J}_0$, subject to $\tilde{B}$ of the emergent magnetic field. The dual filling fraction is therefore given by 
\bea
2\;\tilde{\nu}_{\pm}&=&\frac{\tilde{J}_0}{\tilde{B}/2\pi}=-\frac{B/2\pi}{J_0} \nonumber\\
&=& -\frac{1}{2\;\nu_{\pm}}\,.
\eea
The factor of 2 in the mapping of filling fractions is essential to give the correct particle-hole symmetric (with respect to the lowest Landau level of non-relativistic fermions) interpretation of conjugate Jain-sequence pairs near half filling for each (non-relativistic) species. In~\cite{Son:2015xqa}, this was accomplished by doubling the charge of the single composite fermion, and we expect the same ensues if one restricts the allowed flux configurations as in \cite{Metlitski:2015eka, 2015arXiv150505141W}.\footnote{We thank D. Son for discussions on this point.}

\subsection{Dualizing the Coulomb interaction}\label{subsec:Coulomb}

The other element that will play an important role in our duality is the mapping of the Coulomb interaction of theory A into theory B. Before introducing electromagnetism, both theories have a conserved $U(1)_J$ symmetry that is identified by the duality. This global symmetry can then be weakly gauged by introducing the electromagnetic field $A_\mu$ and adding a 3+1-dimensional kinetic term for this field. In particular, $A_0$ will mediate a Coulomb interaction that can be determined on both sides of the duality. Incidentally, here we consider the canonical kinetic term $\int d^4x F_{\mu\nu}^2$ that leads to a long range Coulomb force, but other types of forces with varying range can be obtained by modifying the action for $A_\mu$.

In theory A, the Coulomb interaction between charge-densities is induced by integrating out the fluctuating part of the $U(1)_{\rm EM} \equiv U(1)_J$ field $A_0$:
\be
S^{\text{(A)}}\supseteq  \frac{1}{2e^2}\int d\omega\;d^3k\; A_0(\omega,k)\;|\vec k|^2\;A_0(-\omega,-k) -\int d\omega\;d^2k \,J_0(\omega,k)\,\int dk_3\,A_0(-\omega,-k) \,,
\ee
where $J_0=\Psi_+^\dag \Psi_+-\Psi_-^\dag \Psi_-$.\footnote{Recall that the matter current has support only on a two-dimensional plane, so that $J_\mu$ depends on $(k_1, k_2)$ but not on $k_3$. In contrast, the electromagnetic field also propagates along $x_3$. The integral over $k_3$ in the last term is the Fourier-transform of the delta-function interaction $\delta(x_3) J_\mu(t, x_1, x_2) A^\mu(t,x_1,x_2,x_3)$.}

Integrating out $A_0$ gives the Coulomb interaction,
\be\label{eq: culmb}
S_{\text{Coulomb}} =-\frac{\pi}{2}\int d\omega\;d^2k\;J_0(\omega,k)\frac{e^2}{|\vec k|}\;J_0(-\omega,-k),
\ee
in the static limit.
This contains both repulsive and attractive terms since $\Psi_+$ and $\Psi_-$ carry opposite charges under the electromagnetic gauge field. Notice that  the interaction is proportional to $1/|\vec k|$ instead of $1/|\vec k|^2$ because $A_0$ and $J_0$ have kinematics in a different number of spatial dimensions. (\ref{eq: culmb}) can be obtained by computing the Coulomb interaction in position space $L_{\text{Coulomb}}\sim -J_0(t,x)\frac{e^2}{|x-y|}J_0(t,y)$, restricting $x$ and $y$ to two spatial dimensions and Fourier-transforming back to 2+1 D.  
$S_{\rm Coulomb}$ is enhanced for small momentum transfer $|\vec k| \rightarrow 0$.

Now we use mirror symmetry to map $J_0$ across the duality. The Coulomb interaction is translated by writing $J_0$, which gives the topological charge in theory B, in terms of the emergent gauge field: 
\be
J_0(x)=\frac{1}{2\pi}\epsilon^{ij}\partial_i a_j(x),\;\; J_0(\omega, k)=\frac{1}{2\pi}\,i\,\epsilon^{ij}\,k_i a_j(\omega, k)\,,
\ee
where we have written the Fourier transformed field in the second equality.

The Coulomb interaction therefore appears as a kinetic term for the spatial components of the emergent gauge field:
\be\label{eq:Coulomb1}
S_{\text{Coulomb}} = -\frac{e^2}{8\pi}\int d\omega\;d^2k\;a_j(\omega,k)\frac{\epsilon^{ij}k_ik_l\epsilon^{lm}}{|\vec k|} a_m(-\omega,-k)\,.
\ee
In this first analysis we are working in the limit $g^2\to \infty$, so we may neglect the gauge field kinetic term compared to (\ref{eq:Coulomb1}). We then integrate out the spatial components $a_i$ using $S_{\text{Coulomb}}$, which will in turn generate a current-current interaction in theory B. To do this, we diagonalize the kinetic matrix by decomposing $a_i$ into longitudinal ($a_L$) and transverse ($a_T$) components:
\be\label{eq:aidecomp}
a_i(\omega,k) =-i\hat k_i a_L(\omega, k)-i \hat k_j \epsilon_{ji} a_T(\omega,k)\,,
\ee
where
\be\label{eq:acurl}
a_L(\omega, k) =i \hat k_j a_j(\omega, k)\;,\;a_T(\omega,k)=i\hat k_j \epsilon_{ji} a_i(\omega, k)\,.
\ee
$\hat k_i$ is the unit vector in momentum space, $\hat k_i = k_i/|\vec k\,|$.

We can then express the pertinent terms in the action of theory B using this basis:
\begin{align}
\label{currentandgauge}
S^{\rm (B)}&\supseteq  -\int_{\omega, k} \tilde{J}^i(\omega,k)a_i(-\omega,-k)+\frac{e^2}{8\pi}\left[\epsilon^{ij}k_ia_j(\omega,k)\right]\frac{1}{|\vec k|} \left[\epsilon^{lm}k_la_m(-\omega,-k)\right]\nonumber\\
&=-\int_{\omega, k}\; \tilde{J}_T(\omega,k)a_T(-\omega,-k)+\tilde{J}_L(\omega,k)a_L(-\omega,-k)+\frac{e^2}{8\pi}a_T(\omega,k)|\vec k|a_T(-\omega,-k),
\end{align}
with $\tilde{J}^i =  \bar{\psi}_+\gamma^i\psi_+-\bar{\psi}_-\gamma^i\psi_-$, the current of the $U(1)$ interaction mediated by $a_i$. We see that only $a_T$ obtains kinematics from the Coulomb interaction. Ignoring the classical kinetic term for the gauge field, as we discussed before, we integrate out $a_T$ and obtain the current-current interaction in theory B:
\be
\label{coulombinB}
S^\text{\rm (B)}_\text{int}=\frac{2 \pi}{e^2}\;\int d\omega d^2k\;\tilde{J}_T(\omega,k)\frac{1}{|\vec k|}\tilde{J}_T(-\omega,-k)\,,
\ee
where $\tilde{J}_T$ is given by the expression,
\bea
\label{currenttransverse}
&&\tilde{J}_T(\omega,k)=i\epsilon^{ij}\hat{k}_i \tilde{J}_j(\omega, k)\nonumber\\
&&=i\epsilon^{ij}\hat{k}_i\sum_{\pm}\,q_\pm \int dp_0\; d^2p\; \bar{\psi}_\pm(p_0,p)\gamma_j \psi_\pm(p_0+\omega,p+k)\,,
\eea
with $q_{\pm} = \pm 1$.

This finishes the derivation of the dual Coulomb interaction in theory B in the limit of $g^2\gg 1$. We note the $e\to 1/e$ mapping of the electric interaction strength between theory A and theory B. This strong/weak duality will be used in \S \ref{sec:dynamics} to derive a perturbatively controlled dual of the quantum Hall system with strong Coulomb interactions.

\subsection{Quantum corrections to scalar fields}

At the classical level, theory B contains fundamental scalars $\phi_{ij}$ (parametrizing the Coulomb branch) and hypermultiplet scalars $u_\pm$. Scalar masses are not generated in the supersymmetric theory, but since we have broken SUSY by turning on the magnetic field (the finite density in theory B), we expect that quantum effects will induce nonzero masses.

It is important to check if these quantum masses are positive definite, to ensure that the theory is stable. In Appendix \ref{app:scalars} we calculate the one loop corrections to $u_\pm$ in the presence of the $\psi_\pm$ Fermi surfaces, obtaining
\be\label{eq:mu}
m_u^2 \propto g^2 k_F >0\,.
\ee
Therefore, the hypermultiplet scalars become massive at one loop and can be safely integrated out at low energies. The fact that $m_u^2>0$ can be understood intuitively as follows. In the low energy theory $E \ll k_F$, the Yukawa interaction $u^\dag \bar \lambda \psi$ is suppressed by powers of $E/k_F$, because the gauginos have momenta around the origin, while for the fermions $p \sim k_F$. In the effective theory near the Fermi surface that will be developed in more detail below, the $u$ fields retain their interactions with the other scalars (such as $|u|^2 \phi^2$) but not with the fermions. The Coleman-Weinberg potential from loops of scalars induces a positive mass squared, which is proportional to the interaction strength times the cutoff of the EFT. This reproduces (\ref{eq:mu}).

We could also incorporate quantum corrections to the masses of Coulomb branch scalars, but instead we will simply turn on 
soft supersymmetry breaking masses to lift the $\phi_{ij}$. This can be done consistently with the duality because the explicit map between the Coulomb branch of theory B and the Higgs branch of theory A is known; see (\ref{eq:quantum-mapb}).\footnote{Note that we cannot do the same for the hypermultiplet scalars, because we cannot map such deformations into theory A.} Since $|\phi| \sim |v|^2$, supersymmetry breaking masses for $\phi$ map to quartic interactions for the scalars in theory A. This deformation is innocuous, because the scalars in theory A were already lifted by the magnetic field, and were not of interest for our purpose. To summarize, by deforming our starting theory A with additional quartic interactions for $v_\pm$, the emergent Coulomb branch scalars become massive and can be integrated out from the low energy theory.

When we construct the low energy theory near the Fermi surface we will then ignore the scalar fields in theory B, and only keep the interactions with the gauge field (whose masslessness is ensured by gauge invariance). As we just discussed, we will also neglect the $k_F$--suppressed interactions with the relativistic gauginos.

\section{Low energy dynamics}\label{sec:dynamics}

In this section we study the low energy dynamics of the dual Fermi surface of composite fermions interacting with the emergent gauge field. 
We do not consider the possible effects of the gauginos.
We set up a renormalization group approach and show that the emergent gauge field flows
from a dynamical exponent $z=3$ at one loop and with $e^2 \ll 1 $ to $z=2$ scaling in the limit of $e\gg 1$. In this new scaling regime, the gauge interaction is classically marginal, enabling us to set up a controlled perturbative expansion in $1/e^2$. This will provide a useful weakly coupled dual of the original strongly coupled quantum Hall system. 
We will then study the induced interaction mediated by the emergent gauge field.
We find a repulsive BCS interaction in all angular momentum channels indicating the stability to superconducting pairing.
Additional interactions are necessary for attraction.

\subsection{Dynamics near the Fermi surface}\label{subsec:compositeFS}

To proceed, let us restrict to a low energy theory of light excitations near the Fermi surface for the composite fermions of theory B. The Appendix shows how to project the original Dirac fermions $\psi_\pm$ in order to keep only the low energy degrees of freedom, to the effect that for $\psi_+$ ($\psi_-$) only the antiparticles (resp. particles) remain. We also review there the one loop renormalization of the gauge field kinetic term due to loops of particles and holes near the Fermi surface and define in more detail the renormalization group scaling.

We work in Landau gauge, $\partial_\mu a_\mu=0$, and in this section only we adopt Euclidean signature for clarity.
The effective action for the light excitations about the Fermi surface (with $B > 0$) takes the form
\be\label{eq:eftB2}
S= S_a+S_f+S_\text{int}\,,
\ee
where the kinetic terms are
\bea
S_a&=&\int d\tau d^2x\,\frac{1}{4g^2} f_{\mu\nu}^2\\
S_f&=& \int dp_0 dp_\perp d\theta\left\{\chi_-^*(p)(ip_0-|p_\perp|)\chi_-(p)+\xi_+^*(-p)(ip_0-|p_\perp|)\xi_+(-p) \right\}\,.
\eea
Here $\xi_+^*$ is the antiparticle component of $\psi_+$, and $\chi_-$ is the particle component of $\psi_-$; see (\ref{fouriertrans2}).
We work in a spherical RG towards the Fermi surface, decomposing the fermion momentum as $\vec p= (k_F+p_\perp)(\cos\theta,\sin\theta)$. The dispersion relation is independent of $\theta$ at leading order in $1/k_F$, so $\theta$ acts as a flavor index (omitted in $\chi$ and $\xi$). See~\cite{shankar, Polchinski:1992ed} for more details.

The interaction terms between the emergent gauge field and the low energy excitations may be deduced from the expression for the current $\t J_\mu$ in (\ref{eq:currents}):
\bea
S_\text{int}&=& -i \int_{p_1,p_2} \left\{ \frac{1}{2}\left(1+e^{- i (\theta(p_1)-\theta(p_2))} \right) a_0(p_2-p_1)+e^{-\frac{i}{2}(\theta(p_1)-\theta(p_2))}a_T(p_2-p_1)\right\}\times \nonumber\\
&&\qquad\times\left(\chi_-^*(p_2)\chi_-(p_1) +\xi_+^*(-p_1)\xi_+(-p_2)\right)
\eea
where $\int_p \equiv \int d^3p/(2\pi)^3$. Importantly, both light excitations have the same charge under $a_\mu$. Also, the interaction with the longitudinal component vanishes because $\t J_L=0$.

These are the classical terms in the action; in order to determine the low energy dynamics, it is necessary to take into account the quantum corrections to the gauge boson propagator. There are three contributions to the gauge boson kinetic terms, which we will now study in more detail: the classical term $\frac{1}{g^2} f_{\mu\nu}^2$, Landau damping and Debye screening from the finite density of composite fermions, and 
the quadratic term (\ref{eq:Coulomb1}) induced by the Coulomb interaction. Let us discuss first the effects of Landau damping and screening, and afterwards consider the Coulomb piece.

As reviewed in Appendix \ref{app:Landau}, the gauge field splits into electric and magnetic components, with inverse propagators
\be\label{eq:Doneloop}
D_\text{el, mag}^{-1}(k)= k^2+ \Pi_\text{el, mag}(k)\,.
\ee
At one loop and for $|k_0| \ll |\vec k|$, the renormalization effects from the Fermi surface screen the electric component,
\be
\Pi_\text{el}(k) \approx M_D^2
\ee
where the Debye mass $M_D^2 \propto k_F$. See (\ref{eq:Pielmag}). Therefore, at low scales this component becomes massive and will be neglected. On the other hand, for the magnetic gauge boson we have
\be\label{eq:Pimag}
\Pi_\text{mag}(k) \approx M_D^2 \frac{|k_0|}{|\vec k|}\,.
\ee
Combining this result with the tree-level kinetic term gives rise to the well-known dynamical exponent $z=3$ for the gauge boson.

Let us now include the effects of the Coulomb interaction. For this, we need to combine (\ref{eq:Doneloop}) with
(\ref{eq:Coulomb1}). The electric component is a combination of $a_0$ and $a_L$; at energy scales much smaller than the Debye mass, this field becomes massive and can be neglected. On the other hand, for the magnetic component we obtain
\be
a_0^\text{mag}=0\;,\;a_i^\text{mag}(k)= i\hat k_n \epsilon_{ni} a_T(k)\,.
\ee
In other words, the magnetic component is equivalent to a scalar field $a_T$. Therefore, the kinetic Lagrangian capturing the tree level term plus Landau damping plus the Coulomb interaction reads
\bea
L_\text{mag}&=&\frac{1}{2}\,a^\text{mag}_i(k) \left((\frac{k^2}{g^2}+\Pi_\text{mag}(k))\delta_{ij}+ \frac{e^2}{4\pi}|\vec k|^{-1}\epsilon_{im} k_m \epsilon_{jn}k_n\right)a^\text{mag}_j(-k)\nonumber\\
&=& \frac{1}{2}\,a_T(k) \left(\frac{k^2}{g^2}+\Pi_\text{mag}(k)+\frac{e^2}{4\pi} |\vec k\,|\right)a_T(-k)\,.
\eea

Plugging in (\ref{eq:Pimag}), we obtain a nontrivial flow for the dynamical exponent. At high momenta $|\vec k| \gg e^2 g^2$, the Coulomb piece is subdominant and the gauge boson has $z=3$ scaling (or $z=1$ if the frequency is sufficiently large). However, at low momenta $|\vec k| \ll e^2 g^2$, the Coulomb piece dominates over the classical $k^2/g^2$ term.
As a result, at low energies we find a $z=2$ dispersion relation and a propagator independent of $g^2$:
\be
D_T^{-1}(k_0, k) \approx M_D^2\,\frac{|k_0|}{|\vec k|} + \frac{e^2}{4 \pi} |\vec k|\,.
\ee
(Given the relation, $J_0 \sim a_T$, the positive sign in the second term results from noting that $J_0$ becomes $i J_0$ when written in Euclidean signature.)
The propagation is now dominated by the interplay between Landau damping and the effective Coulomb contribution. This will have crucial consequences on the low energy dynamics.

In summary, the low energy effective action near the Fermi surface and in the $z=2$ scaling regime for the gauge boson is
\bea\label{eq:eftfinal}
S&=&\frac{1}{2} \int d^3k\,a(k) \left(\t M_D^2\,\frac{|k_0|}{|\vec k|} +|\vec k| \right) a(-k)\\
&+&  \int dp_0 dp_\perp d\theta\,\left\{\chi_-^*(p)(ip_0-|p_\perp|)\chi_-(p)+\xi_+^*(-p)(ip_0-|p_\perp|)\xi_+(-p) \right\} \nonumber\\
&+&i \int dp_0 dp_\perp d\theta\,dp'_0 dp'_\perp d\theta'\,\tilde g(\theta'-\theta) \,a(p-p')\left(\chi_-^*(p)\chi_-(p') +\xi_+^*(-p')\xi_+(-p)\right)\nonumber\,.
\eea
Here we have redefined the boson field,
\be
a(k) \equiv \frac{1}{\t e}\,a_T(k)\;,\;\t M_D^2\equiv \t e^2 M_D^2
\ee
in terms of the new coupling
\be
\tilde e \equiv  \sqrt\frac{4\pi}{e^2}\,.
\ee
Furthermore, the angle-dependent cubic coupling is
\be\label{eq:tg}
\tilde g(\theta'-\theta) \equiv \t e\exp[-i(\theta'-\theta)]\,.
\ee

In terms of this $z=2$ gauge boson, we have an expansion in powers of $\t e$, since the $g$ dependence disappears. Furthermore, combining the $z=2$ scaling with the fermionic RG towards the Fermi surface shows that $\t g$ is classically marginal; see Appendix \ref{app:scaling}. This is very different from what happens with un-deformed mirror symmetry at zero density in theory B or, equivalently, vanishing magnetic field in theory A: the relativistic gauge coupling is classically relevant and quickly leads theory B into a strongly coupled phase, whose weakly coupled dual is theory A. Instead, now the cubic coupling has become classically marginal and hence in the limit of weak $\tilde e$ (i.e., for strong Coulomb interactions in theory A), we have a perturbatively controlled theory. Therefore, there is a range of parameters where theory B is actually the weakly coupled description in the IR. This is a strong/weak duality with respect to the electromagnetic coupling $1/e$.

We note that the emergence of a marginal interaction is reminiscent of the effects of an unscreened Coulomb interaction observed in earlier treatments of the composite fermion approach to the half-filled Landau level of non-relativistic fermions and related systems \cite{halperin1993, nayakwilczekNFL1,NayakWilczekNFL2, Altshuler1994, Mross2010, Metlitski2010}.

\subsection{Superconductivity of composite fermions}\label{subsec:SCcomp}

We are now ready to study the 4-Fermi interactions induced by the interaction with the Coulomb field. 
At low-energies, these will be dominated by the BCS channel; attractive interactions grow towards the IR and cause a superconducting instability~\cite{shankar}. 
We will find the BCS interaction to be repulsive in all angular momentum channels.
Therefore, additional physical ingredients are necessary for a BCS instability.

Following our previous approach, we proceed in two steps. First we discuss the physics when $g^2\to \infty$, ignoring the boson kinetic term and focusing only on the Coulomb interaction. This will allow us to identify explicitly the possible 4-Fermi channels. We will then use the effective field theory (\ref{eq:eftfinal}) to perform a renormalization group analysis of the BCS interactions, including quantum effects from the emergent gauge boson in a Wilsonian way.

\subsubsection{BCS interaction}

In the limit $g^2 \to \infty$, it is sufficient to evaluate the current-current interaction (\ref{coulombinB}) between the excitations near the Fermi surface. 
As discussed previously, when a magnetic field $B>0$ is turned on in theory A, the light excitations of theory B come from antiparticles $\xi_+^*$ of $\psi_+$ and particles $\chi_-$ of $\psi_-$. 
As discussed in Appendix \ref{app:FS}, we may write the current (\ref{currenttransverse}) in terms of the projected low-energy fields,
\be\label{eq:Jcurlfinal}
\t J_T(k)= i\int_{p_1,p_2} \delta^{(3)}(k+p_2-p_1)\, e^{-\frac{i}{2}(\theta(p_1)-\theta(p_2))}\,
 \left(\chi_-^*(p_2)\chi_-(p_1) +\xi_+^*(-p_1)\xi_+(-p_2)\right)\,.
\ee
With this result, we are ready to evaluate the 4-Fermi interaction in Eq. (\ref{coulombinB}) near the Fermi surface:
\bea\label{eq:4fermiint}
S_\text{int}^{(B)}&=&-\frac{2\pi}{e^2}\int_{p_i,p_i'} \delta^3(p_1+p_1'-p_2-p_2')\, \frac{\exp\left[-\frac{i}{2}\left(\theta(p_1)+\theta(p_1')-\theta(p_2)-\theta(p_2')\right)\right]}{|\vec p_1-\vec p_2|}\nonumber\\
&\times& \left(\chi_-^*(p_2)\chi_-(p_1) +\xi_+^*(-p_1)\xi_+(-p_2)\right)\left(\chi_-^*(p_2')\chi_-(p_1') +\xi_+^*(-p_1')\xi_+(-p_2')\right)\,.
\eea

There are three possible BCS pairings: $\langle \chi_- \chi_-\rangle$, $\langle \xi_+ \xi_+\rangle$ (intra-species terms), and $\langle \chi_- \xi_+\rangle$ (i.e., an inter-species coupling). For the first two, the BCS channel sets $\vec p_1\,'=-\vec p_1$, $\vec p_2\,'=\vec p_2$, and the overall angular dependence is proportional to $e^{-i(\theta(p_1)-\theta(p_2))}$. For inter-species pairing, $\vec p_1\,'=\vec p_2$, $\vec p_2\,'=\vec p_1$, and the angular dependence in the exponential factor cancels out. Therefore, the BCS potential becomes
\bea\label{eq:VBCS}
V_\text{BCS}&=&\frac{\pi}{k_Fe^2} \frac{1}{|\sin\frac{\theta(p_1)-\theta(p_2)}{2}|} \Big[ e^{-i(\theta(p_1)-\theta(p_2))}\left( \chi_-^*(p_2)\chi_-^*(-p_2)\chi_-(-p_1) \chi_-(p_1)+(\chi_- \leftrightarrow \xi_+)\right)+\nonumber\\
&+& \left(\chi_-^*(p_2)\xi_+^*(-p_2) \xi_+(-p_1) \chi_-(p_1)+(p_1\leftrightarrow p_2)\right)\Big]\,.
\eea
This indicates that the leading intra-species interaction in the $l=1$ channel is repulsive. 
Likewise, the leading $l=0$ inter-species interaction is repulsive. Repulsive behavior is also found for other angular momentum modes activated by $|\sin(\Delta \theta/2)|$ in (\ref{eq:VBCS}). The same conclusion was reached in Ref. \cite{2014arXiv1403.3694M} for a similar system.\footnote{We thank D. Son for discussions on these 4-Fermi interactions.}

\subsubsection{Renormalization group treatment}

Let us now take into account the full dynamics of the $z=2$ boson, performing an RG treatment of the BCS interaction. We expect this to be useful and nontrivial because the boson-fermion coupling is classically marginal in our system. The renormalization of a Fermi surface coupled to a gapless boson through a nearly marginal interaction was recently addressed in~\cite{Fitzpatrick:2014cfa}, and applied to the BCS coupling in~\cite{Fitzpatrick:2014efa}; we will adopt the same framework here. See~\cite{Son:1998uk, 2014arXiv1403.3694M} for a related approach.

The key point is that integrating over momentum shells, the exchange of gapless bosons leads to a tree level contribution to the BCS beta function. Indeed, integrating this exchange over the tangential direction results in a contribution to the 4-Fermi interaction,
\be
\delta \lambda \propto \t e^2\int dk_\parallel \frac{1}{|k_\parallel|+\t M_D^2 |k_0/k_\parallel|} \sim  \t e^2 \log \frac{\Lambda^2}{\t M_D^2 |k_0|}\,,
\ee
where $\Lambda$ is the UV cutoff. The integral over $k_\parallel$ appears when we change to the angular momentum basis for the 4-Fermi interactions $\lambda_\ell$. Changes in $\Lambda$ can then be absorbed as tree-level contributions to $\lambda_\ell$.

The resulting one loop beta function is
\be\label{eq:betaBCS}
\mu \frac{d\lambda_\ell}{d\mu}= f_\ell\, \t e^2+\frac{\lambda_\ell^2}{2\pi^2}\,,
\ee
where $f_\ell$ is a constant. In our case, $f_{\ell=0}<0$ for all modes. As a result, the interaction mediated by exchange of the massless gauge field provides a constant source term at all energy scales that tends to increase repulsive interactions. In particular, an attractive
BCS interaction weakens as the length scale is increased and there are no superconducting instabilities at this order. One intriguing feature of (\ref{eq:betaBCS}) is the existence of a UV fixed point (i.e., an IR unstable fixed point) at
\be
\lambda_\ell^*=\sqrt{2\pi^2 |f_\ell|\t e^2}\,,
\ee
which is perturbatively controlled for small $\t e$. If this fixed point is not lifted by higher order quantum corrections, it would be important to understand its dual interpretation in the quantum Hall system of theory A.

It is interesting to ask what additional interactions must be added to make the BCS interaction attractive.
If the interactions can be made attractive, we expect an enhancement of the pairing scale similar to that which occurs in QCD at finite density due to exchange of magnetic gluons~\cite{Son:1998uk}. 
Pairing of (non-relativistic) composite fermions is believed to result in either the Moore-Read Pfaffian state or its particle-hole conjugate, the Anti-Pfaffian state \cite{lee2007antipf, levin2007}.
The former occurs in the $\ell=1$ angular momentum channel, while the latter obtains in the $\ell=-3$ channel. While our current analysis cannot determine the ground state of the theory, it may provide a controlled framework to understand what interactions in theory A lead to attractive BCS interactions in theory B in the requisite channels.

\section{Conclusions and future directions}

In this paper, we have demonstrated how  mirror symmetry can be applied to study the physics of 2+1D Dirac fermions in a finite transverse magnetic field at filling fractions $\nu \sim 0$ for each species, i.e., about the half-filled zeroth Landau level of the Dirac fermions.
We derived a dual description in terms of electrically neutral composite fermions at finite density, interacting via an emergent gauge field.
This dual description is similar, appropriately modified to the Dirac context, to that which has been advocated by a number of authors \cite{shankarmurthyshort, Lee1998, PasquierHaldane, Read1998, simon1998, Stern1999, MurthyShankar2003} for the description of the half-filled Landau level of non-relativistic fermions.
(See Refs. \cite{alicea2005, Mross2015} for related instances.)
Mirror symmetry provides a concrete derivation of flux attachment and allows us to understand precisely the interactions between composite fermions in the emergent non-Fermi liquid.

Mirror symmetry has the following remarkable implication: the ground state wave function (singled out by residual interactions at finite $g$ or Coulomb interactions) of the half-filled zeroth Landau level of Dirac fermions, theory A, describes the non-Fermi liquid ground state of a finite density of Dirac fermions interacting through an emergent gauge field, theory B.\footnote{We thank S. Raghu for related discussions.}
In the limit of zero Landau level mixing, vanishing fluctuating Coulomb field, and neglecting any additional irrelevant interactions, i.e., in the limit where the low-energy theory is non-interacting, we may easily solve for the theory A reduced density matrix for some spatial subregion.
The implication is that this reduced density matrix is related to the reduced density matrix of a particular non-Fermi liquid.
For instance, we might naively anticipate a logarithmic violation of the entanglement entropy area law of the theory B (non-)Fermi liquid to be visible via a simpler calculation in theory A.
The equality of 3-sphere partition functions of theory A and theory B (at vanishing external field in theory A and vanishing fermion density in theory B) \cite{KapustinWillettYaakov} implies the equality of the constant subleading term in the entanglement entropy of the ground states of the two theories for a disk subregion (when such perturbations are not present).

In the future, we would like to extend the duality to the study of non-relativistic, rather than Dirac, fermions.
We expect to obtain a non-relativistic theory from our Dirac starting point in theory A by turning on a finite chemical potential. 
The addition of mass terms for either of the $\psi_{\pm}$ fermions of theory B results in a non-zero (level $1/2$) Chern-Simons term for the emergent gauge field and may bring us intriguingly close to the composite Fermi liquid theory of Refs. \cite{Kalmeyer1992, halperin1993}, except that the composite fermions carry zero electromagnetic charge. Furthermore, the nearly marginal coupling of the $z=2$ gauge boson to the composite fermions of theory B suggests the remarkable possibility of a controlled quantum critical metal, something that we hope to analyze in future work.

In the course of finishing this paper, we became aware of the recent works in Refs. \cite{Metlitski:2015eka, 2015arXiv150505141W} that study duality of a single Dirac fermion that arises on the bounding surface of a time-reversal invariant topological insulator.
We anticipate that the duality involving a single Dirac cone can be understood from mirror symmetry by realizing theory A on a domain wall in 3+1 dimensions and then separating the two Dirac cones spatially by a symmetry-breaking perturbation.

Finally, we remark that the structure of theory B shares an additional similarity with the theory proposed to describe a putative second order transition between the composite Fermi liquid of Refs. \cite{Kalmeyer1992, halperin1993} and the anti-composite Fermi liquid introduced in \cite{BMF2015}.
The anti-composite Fermi liquid was introduced in response to the surprising experiments \cite{kamburov2014, LiuDengWaltz} that observe magnetoresistance oscillations that imply a (composite) Fermi wave vector tied to the electron/hole density for $\nu_{\rm NR}<1/2$ or $\nu_{\rm NR} > 1/2$, respectively.
It is curious that two flavors of fermions naturally arise out of the simplest example provided by mirror symmetry.

\section*{Acknowledgments}
S.K. thanks D.T. Son for important communications which motivated this project, and the Simons
Symposium in San Juan, Puerto Rico for providing a very stimulating environment for related discussions.  We are also grateful to S. Raghu for educational discussions about many aspects
of this and related problems.
We thank E. Fradkin and D.T. Son for comments on an early draft of this work. 
S.K. is supported by the NSF via grant PHY-0756174 and the DoE
Office of Basic Energy Sciences contract DE-AC02-76SF00515.  S.K. and M.M. also acknowledge the
support of the John Templeton Foundation.  G.T. is supported by CONICET, and PIP grant 11220110100752. H.W.~ is supported by a Stanford
Graduate Fellowship.

\appendix

\section{Relativistic fermions at finite density}

In this Appendix we discuss the low energy theory for massless Dirac fermions at finite density, in 2+1 dimensions.  We also review the one loop renormalization of the gauge field and discuss the RG scaling of the coupled fermion--boson system.

\subsection{Light quasiparticles near the Fermi surface}\label{app:FS}

The classical Lagrangian is
\be
L = \bar \psi(i\gamma^\mu \partial_\mu + \mu_F \gamma^0) \psi
\ee
so that the equation of motion requires
\be\label{eq:mass-shell}
p_0= - \mu_F \pm |\vec p\,|\,.
\ee
This describes particles (the plus sign above) and antiparticles (the minus sign).

For $\mu_F<0$, the antiparticles can have low energy for $|\vec p\,| \sim \mu_F$, while the particles have high energies $p_0 \sim - \mu_F$. 
This is the situation for the $\psi_+$ fermions of theory B when a magnetic field $B > 0$ in turned on in theory A.
On the other hand, for $\mu_F>0$ it is the particles that are light and the antiparticles that are heavy. 
This is the situation for the $\psi_-$ fermions of theory B when the magnetic field $B > 0$.

We now describe how to construct a low energy theory for the excitations near the Fermi surface by keeping only the light excitations and projecting out the heavy modes.
For this, we note that the Dirac equation $\left((p_0+\mu_F) \gamma^0+p_i \gamma^i\right) \psi(p)=0$ requires
\be
\left(1\pm \gamma^0 \frac{\gamma^i p_i}{|\vec p\,|} \right) \psi(p) =0
\ee
on the mass shell (\ref{eq:mass-shell}). 
Let us then define the projectors
\be\label{eq:P1}
P^{(\pm)}(p) \equiv \frac{1}{2}\left(1\pm \gamma^0 \frac{\gamma^i p_i}{|\vec p\,|} \right)\,,
\ee
in terms of which the Dirac Lagrangian becomes
\be
L = \psi^\dag \Big((p_0+\mu_F -|\vec p\,|)P^{(-)}(p)+(p_0+\mu_F +|\vec p\,|)P^{(+)}(p) \Big) \psi\,.
\ee
Here the first term proportional to $P^{(-)}$ projects out antiparticles (i.e. it keeps the particles only), while $P^{(+)}$ projects out particles (keeps antiparticles).
For the representation of gamma matrices used in the paper, these projectors become
\be\label{eq:P2}
P^{(\pm)}(p)=\frac{1}{2} \left(\begin{matrix} 1 && \mp i e^{- i\theta} \\ \pm i e^{i\theta}&&1\end{matrix}\right)
\ee
where $\vec p = p^i=|\vec p\,|(\cos \theta,\sin\theta)$. (Recall that $p_i=-p^i$ in our metric signature.)

We expand $\psi$ in terms of its particle, $\chi$, and antiparticle, $\xi^\ast$, components:
\begin{align}
\label{fouriertrans}
\psi(x) = \int dp_0 d^2p\ \Big({1 \over \sqrt{2}} \begin{pmatrix}
i e^{- i \theta} \cr 1
\end{pmatrix}
\chi(p)\,e^{-i p\cdot x}
+
{1 \over \sqrt{2}} \begin{pmatrix}
-i e^{- i \theta} \cr 1
\end{pmatrix}
\xi^\ast(p)\,\,e^{i p \cdot x}\Big),
\end{align}
so that 
\begin{align}
P^{(-)}(p) \psi(p) = {1 \over \sqrt{2}} \begin{pmatrix}
 i e^{- i \theta} \cr 1
\end{pmatrix}
\chi(p), \quad P^{(+)}(p) \psi(p) = {1 \over \sqrt{2}} \begin{pmatrix}
-i e^{- i \theta} \cr 1
\end{pmatrix}
\xi^\ast(p).
\end{align}
Notice that $\xi$ is complex-conjugated above, consistent with its interpretation as the antiparticle component of $\psi$. The
Dirac Lagrangian in terms of the particle/antiparticle excitations,
\begin{align}
L & = \chi^\ast \Big(p_0 + \mu_F - |\vec{p}\,| \Big) \chi + \xi \Big(- p_0 + \mu_F + |\vec{p}\,| \Big) \xi^\ast.
\end{align}

Depending on the sign of $\mu_F$, the low energy theory near the Fermi surface will keep one of these two contributions: for $\mu_F >0$ the effective theory will contain particles, while antiparticles are light if $\mu_F<0$. On both sides of the mirror duality, we have two types of fermions $\Psi_\pm$ and $\psi_\pm$, and the sign of the chemical potentials for each species is determined by the sign of the charges and any background fields. This means that there will be a Fermi surface of particles for one type of fermion, and a Fermi surface of antiparticles for the other type. 
In particular, for magnetic field $B>0$ in theory A, the low-energy description near the Fermi surface of theory B consists of antiparticles of $\psi_+$ and particles of $\psi_-$ and is given by:
\begin{align}
\label{eq:eftB0}
L_{\rm eff}^{(B)} = \xi^\ast_+ \Big(p_0  - |p_\perp | \Big) \xi_+ + \chi^\ast_- \Big(p_0 -|p_\perp | \Big) \chi_-+L_\text{gauge}\,,
\end{align}
where we denote the antiparticle component of $\psi_+$ by $\xi_+$ and the particle component of $\psi_-$ by $\chi_-$. We have decomposed the momentum as
\be
\vec p = (k_F+p_\perp) (\cos\theta(p),\, \sin\theta(p))\;,\; k_F=|\mu_F|\,.
\ee
Also, $L_\text{gauge}$ are the interactions with the emergent gauge field, which will be discussed shortly.

In order to determine the interaction between the emergent gauge field and the low energy fermions, let us now calculate the current
\be
\t J^\mu= \bar \psi_+ \gamma^\mu \psi_+- \bar \psi_- \gamma^\mu \psi_-\,.
\ee
for the projected composite fermions.
For this, it is convenient to redefine the antiparticle momentum $p \to -p$ in (\ref{fouriertrans}), which simplifies the expression for the fermion in momentum space,
\begin{align}
\label{fouriertrans2}
\psi(p) =  u(p)\Big(
\chi(p)\,
+
\xi^\ast(-p)\Big)\;,\; u(p) \equiv {1 \over \sqrt{2}} \begin{pmatrix}
i e^{- i \theta(p)} \cr 1
\end{pmatrix}\,.
\end{align}
This can be seen as a consequence of $P^{(+)}(-p) =P^{(-)}(p)$. We will need the following spinor identities:
\bea
&&u^\dag(p_2)P^{(-)}(p_2) P^{(-)}(p_1)u(p_1)=\frac{1}{2}\left(1+e^{- i (\theta(p_1)-\theta(p_2))} \right)\\
&&u^\dag(p_2)P^{(-)}(p_2) (\gamma^0  \epsilon_{ij}\hat k^i \gamma^j)P^{(-)}(p_1)u(p_1)= e^{-\frac{i}{2}(\theta(p_1)-\theta(p_2))}\,\sin\left(\frac{1}{2}\theta(p_1)+\frac{1}{2}\theta(p_2)-\theta(p_1-p_2)\right)\nonumber\\
&&u^\dag(p_2)P^{(-)}(p_2) (\gamma^0 \hat k^i \gamma^i)P^{(-)}(p_1)u(p_1)=  e^{-\frac{i}{2}(\theta(p_1)-\theta(p_2))}\,\cos\left(\frac{1}{2}\theta(p_1)+\frac{1}{2}\theta(p_2)-\theta(p_1-p_2)\right)\nonumber
\eea
where $k=p_1-p_2$. Noting that
\be
\theta(p_1-p_2)=\frac{\pi}{2}+\frac{\theta(p_1)+\theta(p_2)}{2}\,,
\ee
these simplify to
\bea\label{eq:spinor-identities}
&&u^\dag(p_2)P^{(-)}(p_2) P^{(-)}(p_1)u(p_1)=\frac{1}{2}\left(1+e^{- i (\theta(p_1)-\theta(p_2))} \right)\nonumber\\
&&u^\dag(p_2)P^{(-)}(p_2) (\gamma^0  \epsilon_{ij}\hat k^i \gamma^j)P^{(-)}(p_1)u(p_1)=- e^{-\frac{i}{2}(\theta(p_1)-\theta(p_2))}\\
&&u^\dag(p_2)P^{(-)}(p_2) (\gamma^0 \hat k^i \gamma^i)P^{(-)}(p_1)u(p_1)=  0 \nonumber\,.
\eea

The resulting expressions for $\t J_\mu$ in terms of low energy excitations are
\bea\label{eq:currents}
\t J_0(k)&=& - \int_{p_1,p_2} \delta^{(3)}(k+p_2-p_1)\,\frac{1}{2}\left(1+e^{- i (\theta(p_1)-\theta(p_2))} \right)\left(\chi_-^*(p_2)\chi_-(p_1) +\xi_+^*(-p_1)\xi_+(-p_2)\right), \nonumber\\
\t J_T(k)&=& i\int_{p_1,p_2} \delta^{(3)}(k+p_2-p_1)\, e^{-\frac{i}{2}(\theta(p_1)-\theta(p_2))}\,
 \left(\chi_-^*(p_2)\chi_-(p_1) +\xi_+^*(-p_1)\xi_+(-p_2)\right), \nonumber\\
 \t J_L(k)&=&0,
\eea
where $\int_{p} \equiv \int d^3 p/(2\pi)^3$. Recall the decomposition into longitudinal and transverse components given in (\ref{eq:acurl}).

Given these results for the current, it is straightforward to add the interactions with the emergent gauge field. The resulting Lagrangian for the light fermions in theory B is
\bea\label{eq:eftB}
S_{\rm fermion}^{(B)}&=&\int_p\left\{\chi_-^*(p)(p_0-|p_\perp|)\chi_-(p)+\xi_+^*(-p)(p_0-|p_\perp|)\xi_+(-p)\right\} \nonumber\\
&+& \int_{p_1,p_2} \left\{ \frac{1}{2}\left(1+e^{- i(\theta(p_1)-\theta(p_2))} \right) a_0(p_2-p_1)+ie^{-\frac{i}{2}(\theta(p_1)-\theta(p_2))}a_T(p_2-p_1)\right\}\times \nonumber\\
&&\qquad\times\left(\chi_-^*(p_2)\chi_-(p_1) +\xi_+^*(-p_1)\xi_+(-p_2)\right)\,.
\eea
Note that these low energy excitations carry the same charge under the emergent gauge field $a_\mu$.

\subsection{One loop renormalization of the gauge field propagator}\label{app:Landau}

We will now review the one loop quantum corrections to the gauge field propagator, originating from Fermi surface loops. See e.g., Ref. \cite{Miransky:2001qs} for more details.

For simplicity we fix to Landau gauge $\partial_\mu a_\mu=0$, but the analysis for more general $\xi$ gauges is very similar. We also work in Euclidean signature.

In the presence of a Fermi surface, which breaks Lorentz invariance, the general form of the inverse gauge field propagator including quantum corrections is
\be\label{eq:Dgeneral}
D^{-1}(k)_{\mu\nu}=(k^2+\Pi_\text{mag}(k)) \mathcal O^\text{mag}_{\mu\nu}(k)+(k^2+\Pi_\text{el}(k)) \mathcal O^\text{el}_{\mu\nu}(k)\,,
\ee
where $\mathcal O$ project to the magnetic and electric components of the gauge field,
\bea
\mathcal O^\text{mag}_{\mu\nu}(k)&=& \delta_{\mu\nu}- u_\mu u_\nu-\frac{\vec k_\mu \vec k_\nu}{|\vec k|^2}\nonumber\\
\mathcal O^\text{el}_{\mu\nu}(k)&=&u_\mu u_\nu+\frac{\vec k_\mu \vec k_\nu}{|\vec k|^2}-\frac{k_\mu k_\nu}{k^2}\,.
\eea
Here, $u_\mu=(1,0,0)$, $\vec k_\mu= k_\mu-(u\cdot k) u_\mu$, and $k^2=k_0^2+\vec k^2$. $\Pi_\text{mag, el}$ are the quantum corrections from the finite density fermions. 

At one loop,
\be\label{eq:Pielmag}
\Pi_\text{mag}(k)=M_D^2 \frac{k_0^2}{|\vec k|^2}\left(\sqrt{1+\frac{|\vec k|^2}{k_0^2}}-1\right)\;,\;\Pi_\text{el}(k)=M_D^2-\Pi_\text{mag}(k)\,.
\ee
The projected gauge fields are denoted by $a^\text{mag}_\mu=\mathcal O^\text{mag}_{\mu\nu} a_\nu$ and similarly for the electric component. Note that $\mathcal O^\text{mag}_{\mu\nu}$ only has spatial components.

For completeness, let us also analyze the one loop action for $a_i$, together with the Coulomb induced term, in Minkowski signature $(+--)$. Focusing on the magnetic component, which is the one that survives in the low energy theory, we write its action as
\be
S= \int d^3 k\, a^\text{mag}_i(k) K_{ij}(k) a^\text{mag}_i(-k)\,,
\ee
where the kernel $K_{ij}$ combines the tree level, one loop and Coulomb pieces:
\be
K_{ij}(k)=\frac{1}{2g^2}(k_0^2-|\vec k|^2-\Pi_\text{mag}(k))\delta_{ij}-\frac{e^2}{8\pi}|\vec k|\, \epsilon_{il} \hat k_l \epsilon_{jn} \hat k_n\,,
\ee
and in this signature,
\be
\Pi_\text{mag}(k)=M_D^2 \frac{k_0^2}{|\vec k|^2} \left(1-\sqrt{1-\frac{k_0^2}{|\vec k|^2}}\right)\,.
\ee
The on-shell condition is satisfied by two dispersion relations,
\be
k_0^2=|\vec k|^2+\Pi_\text{mag}(k)
\ee
which is the longitudinal part and is hence orthogonal to $a_i^\text{mag}$, and
\be
k_0^2=|\vec k|^2+\Pi_\text{mag}(k)+\frac{e^2 g^2}{4\pi}|\vec k|
\ee
whose eigenvector is proportional to the transverse component of the field. This is the dispersion relation used for $a_T$ in the main text.

\subsection{Renormalization group scaling}\label{app:scaling}

In this section we discuss the scaling of the fermion-boson system for a theory with general spatial dimension $d$ and boson dynamical exponent $z$:
\bea
S&=&\int dq_0 d^{d-1}q\,a(q) \left(\t M_D^2\,\frac{|q_0|}{|\vec q\,|} +|\vec q\,|^z \right) a(-q)+\int dp_0 dp_\perp d^{d-1}\hat n \,\psi^\dagger(p) (ip_0-p_\perp) \psi(p) \nonumber\\
&+& \int dp_0 dp'_0 dp_\perp dp'_\perp\, d^{d-1}(\hat n+\hat n') d^{d-1}(\hat n-\hat n')\, \tilde g\, a(p-p') \psi^\dagger(p) \psi(p')\,.
\eea
The form of the boson kinetic term is motivated by the term that appears in the main text.

First, the $d$-dimensional fermionic momenta are divided into a direction perpendicular to the Fermi surface, and $d-1$ angles $\hat n$:
\be
\vec p = \hat n (k_F + p_\perp)\,.
\ee
The fermion scaling is then
\be
[p_0]= [p_\perp]=1\;,\;[\psi(p)]=-\frac{3}{2}\,,
\ee
independent of $d$.

Next, from the cubic interaction we recognize that the bosonic momenta relevant for quantum effects are differences of fermionic momenta, $\vec q= \vec p\, -\vec p\,'$. More explicitly,
\be
\vec q= \hat n q_\perp+\vec q_\parallel\;,\; q_\perp= p_\perp-p_\perp'\,,
\ee
and
\be
\vec q_\parallel= k_F (\hat n- \hat n')\,.
\ee
Also, $q_0=p_0-p_0'$. These conditions fix
\be
[q_0]= [q_\perp]=1
\ee
for the boson, and we still need to decide how to scale $q_\parallel$, namely the difference between fermionic angles. 

The scaling of $\hat n -\hat n'$ happens due to a dynamical reason: the boson scatters fermions predominantly in tangential directions to the Fermi surface. At a given energy, this defines the size of the patch around each fermion that will be more strongly coupled due to boson exchange. This tangential scaling is obtained from the boson propagator, neglecting the $q_\perp$ dependence (which can be checked self-consistently). Doing so, we obtain
\be
[q_\parallel]=\frac{1}{z}\;,\;[a(q)]=-\frac{d+3z-2}{2z}\,.
\ee

Finally, plugging these scaling dimensions into the cubic interaction gives
\be
[\t g]=\frac{z-d}{2z}\,.
\ee
In particular, for the values of the model in the main text, $d=2$, $z=2$, we find a classically marginal interaction.

This scaling is related to, but not the same as, the patch scaling used in other works such as \cite{ nayakwilczekNFL1,NayakWilczekNFL2, Altshuler1994, Mross2010}. In particular, the spherical RG for the fermionic sector guarantees the marginality of the BCS interaction in every dimension.

\subsection{One loop corrections to scalars}\label{app:scalars}

Let us evaluate the one loop corrections to theory B scalars. We focus on the hypermultiplet scalars $u_\pm$, lifting the Coulomb branch scalars by explicit soft SUSY breaking masses.

The scalars $u_\pm$ interact with the emergent gauge field, Coulomb branch scalars, with themselves and with the gauginos and hypermultiplet fermions. In the absence of finite density, the perturbative corrections cancel exactly. Therefore, in order to understand quantum corrections due to finite density, it is sufficient to compute the one loop diagrams from the cubic vertices of the schematic form $u \bar \lambda \psi$.

These one loop effects are a bit atypical, in that one fermion line (from the gaugino) is relativistic, while the other (from $\psi_\pm$) is at finite density. Let us then discuss a model of the form
\be
L = |\partial u|^2 - \bar \psi (\not \!\partial - \gamma_0 \mu_F) \psi- \bar \lambda \not \!\partial \lambda-g u^\dag \bar \lambda \psi-g u \bar \psi \lambda\,.
\ee
We work in the euclidean formalism, where the chemical potential appears as an imaginary background $A_0$. We first compute the one loop corrections in this model, and then specialize the result to theory B.

At one loop, the scalar kinetic term changes to $p^2+\Pi(p)$, where
\be
\Pi(p) = g^2 \int\frac{d^3 q}{(2\pi)^3} G_\lambda(q) G_\psi(p+q)= g^2 \int\frac{d^3 q}{(2\pi)^3} \frac{i}{\not \! q} \frac{i}{\not \! q+\not \! p-i \gamma_0 \mu_F}\,.
\ee
In order to find the RG evolution of the mass, it is sufficient to set the external momenta $p=0$ and work with a lower cutoff $\omega \sim p$. The UV cutoff is denoted by $\Lambda$. Then we need to calculate
\be
\Pi=-2g^2 \int_\omega^\Lambda \frac{dq_0}{2\pi} \frac{d^2 q}{(2\pi)^2}\,\frac{q_0(q_0-i\mu_F)+q^2}{[(q_0-i\mu_F)^2+q^2](q_0^2+q^2)}\,,
\ee
where the factor of $2$ comes from the two components of the fermions, and here $q \equiv |\vec q\,|$.

Performing the $q_0$ integral by residues gives, after a few simplifications,
\be
\Pi=-g^2 P \int_\omega^\Lambda \frac{q dq}{2\pi} \left(\frac{1}{2q-\mu_F}+\frac{\Theta(q+\mu_F)}{2q+\mu_F}-\frac{\Theta(\mu_F-q)}{2q-\mu_F} \right)\,,
\ee
and $P$ denotes the principal value.
This shows that as the IR scale $\omega$ crosses $\mu_F$, new contributions are generated due to the finite density.

Let us consider first $\mu_F>0$. We have
\be
\Pi=-g^2 \left\{ P\int_\omega^\Lambda \frac{q dq}{2\pi} \left(\frac{1}{2q-\mu_F}+\frac{1}{2q+\mu_F} \right)-\Theta(\mu_F-\omega)\,P\int_\omega^{\mu_F} \frac{q dq}{2\pi}\frac{1}{2q-\mu_F} \right\}\,.
\ee
Using
\be
P \int_\omega^\Lambda \frac{q dq}{2q-\mu_F}=\frac{1}{2}(\Lambda-\omega)+\frac{\mu_F}{2}\,\log\frac{2\Lambda-\mu}{\mu-2\omega}
\ee
and focusing on low frequencies $\omega \ll \mu_F$, we obtain
\be
\Pi(\omega \to 0) \approx -\frac{g^2}{2\pi} \left(\Lambda-\frac{\mu_F}{2}+\frac{\mu_F}{4}\,\log\frac{2\Lambda-\mu_F}{2\Lambda+\mu_F} \right)+\mathcal O(\omega)\,.
\ee
The term linear in the cutoff $\Lambda \gg \mu_F$ is the well-known (Coleman-Weinberg type) negative contribution to the scalar mass. This piece is cancelled by supersymmetry, i.e., by the contribution from the bosonic fields. On the other hand, at finite density we find an additional contribution proportional to $\mu_F$, with a sign that is opposite from the cutoff piece. Therefore, quantum corrections from finite density tend to stabilize the scalars.
The logarithmic contribution vanishes in the limit $\Lambda \rightarrow \infty$.
Following similar steps, the result for $\mu_F<0$ also shows an additional contribution proportional to $|\mu_F|$ that tends to increase the scalar mass.

In the model for theory B, the scalars $u_\pm$ interact with Fermi surfaces of particles and antiparticles. Adding both effects, we finally obtain
\be
\Pi_u \approx \frac{g^2}{2\pi}|\mu_F|
\ee
showing that in the low energy theory, the hypermultiplet scalars are stabilized at a very high scale and may be safely integrated out.

\bibliography{QHE.bib}{}
\bibliographystyle{utphys}
\end{document}